\documentclass[12pt]{article}
\usepackage{epsfig, amssymb, graphicx, amsmath} 

\usepackage{eurosym}	
\usepackage{wrapfig}	
\usepackage{float}		
\usepackage{subfig}	
\usepackage{footnote}
\usepackage{amsthm} 	 

\usepackage{bbm}		

\usepackage[authoryear]{natbib}
\bibpunct{(}{)}{,}{a}{}{,}

\numberwithin{theorem}{section} 

\theoremstyle{definition}

\theoremstyle{remark}

\newcommand{\be}{\begin{equation}}
\newcommand{\en}{\end{equation}}
\newcommand{\ben}{\begin{equation*}}
\newcommand{\enn}{\end{equation*}}
\newcommand{\bea}{\begin{eqnarray}}
\newcommand{\ena}{\end{eqnarray}}

\DeclareMathOperator*{\argmax}{argmax}

\begin{document}
 
\newlength\tindent
\setlength{\tindent}{\parindent}
\setlength{\parindent}{0pt}
\renewcommand{\indent}{\hspace*{\tindent}}

\begin{savenotes}
\title{
\bf{A closed formula for illiquid corporate bonds \\
and an application to the European market}}
\author{
Roberto Baviera$^\dagger$ \& 
Aldo Nassigh$^\ddagger$ \&
Emanuele Nastasi$^\flat$ 
}

\maketitle

\vspace*{0.11truein}
\begin{tabular}{ll}
$(\dagger)$ &  Politecnico di Milano, Department of Mathematics, 32 p.zza L. da Vinci, 20133 Milano \\
$(\ddagger)$ & Unicredit S.p.A., p.zza  Gae Aulenti, 20154 Milano\footnote{The views expressed here are those of the author and not necessarily those of the bank.} \\
$(\flat)$ & Exprivia S.p.A.,  43 via dei Valtorta, 20127 Milano.
\end{tabular}
\end{savenotes}

\vspace*{0.11truein}

\begin{abstract}
\noindent
We propose an option approach for pricing bond illiquidity that is reminiscent of the celebrated work of Longstaff (1995)  on the non-marketability of some non-dividend-paying shares in IPOs. 
This approach describes a quite common situation in the fixed income market: 
it is rather usual to find issuers that, besides liquid benchmark bonds, issue some other bonds that either are placed to a small number of investors in private placements or have a limited issue size.

\noindent
We model  interest rate and credit risks via a convenient reduced-form approach.
We deduce a simple closed formula for illiquid corporate coupon bond prices when liquid bonds with similar characteristics (e.g. maturity) are present in the market for the same issuer. 
The key model parameter is the time-to-liquidate a position, i.e. the time that an experienced bond trader takes to liquidate a given position on a corporate coupon bond.
We show that illiquid bonds present an additional liquidity spread that depends on the time-to-liquidate aside from bond volatility.

\noindent
We provide a detailed application for two issuers in the European market.
\end{abstract}

\vspace*{0.11truein}
{\bf Keywords}: 
Corporate coupon bonds, liquidity, time-to-liquidate.
\vspace*{0.11truein}

{\bf JEL Classification}: 
C51, 
G12, 
H63. 

\vspace{0.6cm}
\begin{flushleft}
{\bf Address for correspondence:}\\
Roberto Baviera\\
Department of Mathematics \\
Politecnico di Milano\\
32 p.zza Leonardo da Vinci \\ 
I-20133 Milano, Italy \\
Tel. +39-02-2399 4575\\
roberto.baviera@polimi.it
\end{flushleft}

\newpage

\begin{center}
\Large\bfseries 
A closed formula for illiquid corporate bonds \\
and an application to the European market
\end{center}


\vspace*{0.21truein}


\section{Introduction}
\label{sec:introduction}

The natural question that arises when dealing with liquidity is: ``How long does it take to liquidate a given position?''.
Despite the relevance of this question, not only has there not yet appeared, in the financial industry, a unique modeling framework, but not even a standard language for addressing liquidity. 
Unfortunately, liquidity problems are, in general, really complicated. 
There are several aspects of asset liquidity, including tightness (i.e. bid--ask spread, the transaction cost incurred  in case of a small liquidation), 
market impact (i.e. the average response of prices to a trade, see, e.g. \citet{BFL}), 
market elasticity (i.e. how rapidly a market regenerates the liquidity removed by a trade) and the time-to-liquidate a position. 
 
Traditional liquidity measures were developed for the equity market within Market Impact Models \citep[see, e.g.][and references therein]{FLM, BFL, Gatheral}  
with a particular focus on stocks with larger capitalization: 
execution typically takes place in a timeframe from minutes to hours.
However, 
these liquidity measures are not applicable to securities, 
such as many corporate bonds, that do not trade on a regular basis: 
often prices of many illiquid corporate bonds are not observed in the marketplace for several days. 
In this case a complete representation of asset liquidity could be 
not feasible, 
for several reasons, such as
i) the market is still largely OTC and bid--ask quotes are not available for many corporate 
bonds\footnote{Practitioners well know that publicly disclosed quotes are often not true commitments to trade at that price but rather just indications (i.e. `indicative' quotes).}
ii) trading costs often decrease with trade size \citep[see, e.g.][]{Edwards} and
iii) the time-to-liquidate a position can be some weeks, or even months, in some cases. 

\bigskip

A focus on bond market liquidity was stimulated by 
the regulatory effort to introduce more transparency in the bond market. 
In the U.S.A., 
starting from the $1^{st}$ of July, 2002, information on the prices and the volumes of completed transactions have been publicly disclosed for a significant set of corporate bonds. 
The National Association of Security Dealers (NASD, and after July 2007 the Financial Industry Regulatory Authority, FINRA) mandated 
post-trade transparency in the corporate bond market 
through the Trade Reporting and Compliance Engine (TRACE) program; under TRACE, all trades for corporate bonds in USD must be reported within 15 minutes of execution
 \citep[see, e.g.][and references therein]{NFL}.
This dataset has boosted an econometric research on corporate bond 
liquidity \citep[see, e.g.][]{BMV,  NFL, Helwege, Schestag2016, Asquith2013}; econometric studies 
that have clarified several aspects of bond liquidity and
that have stimulated new research questions.

Thanks to the transactional data provided by TRACE,
\citet{NFL}  found in the corporate bond spread  
a significant evidence of
a liquidity component  in addition to the default risk component, thus contributing to explain the so-called ``credit spread puzzle".

Focusing on the most liquid bonds, 
\citet{Schestag2016} 
were able to apply to the bond market eight competing intraday liquidity measures\footnote{Six transaction cost measures, one price impact measure, and one price dispersion measure.}  
and to benchmark the effectiveness of thirteen liquidity proxies that only need daily information.
They provided guidance  on 
the three low frequency proxies that perform better 
when daily data are the only available ones. 
Their analysis suggests a relevant question for practitioners: 
what price can be associated to unquoted bonds, in particular when their prices are not available for several days in the marketplace?

In \citet{NFL} credit and liquidity were commingled, the liquidity proxies depended on the credit quality of the issuer; therefore,
\citet{Helwege} proposed to identify a {\it  sheer liquidity} premium: a component in bond price that depends uniquely on market liquidity.
They measured the difference in the spreads between matched pairs of bonds with the same 
characteristics\footnote{Same issuer, same coupon type and both coupon amount and maturity within a narrow range.} except market liquidity.
They highlighted that 
it is quite difficult to separate empirically the two components of credit and liquidity in corporate bond yields with standard econometric techniques;
once they measured the {\it sheer liquidity} premium, 
they found 
that it was time-varying and  
that it was related to the observed market conditions.  
Their approach appears very fruitful and suggests an interesting line of research: 
identifying the {\it sheer liquidity} component in corporate bond yields could allow 
to pick out the relevant risk factors in the liquidity spread, 
e.g. the volatility of the bond or the time-to-liquidate a given bond position, mentioned in the question we have started with in this Introduction.  


Moreover, the analysis of TRACE database addressed
the impact  on bond market liquidity following the introduction of this post-trade program.
The consequences of this program were mixed, as discussed by 
\citet{Asquith2013}, 
with a decrease in daily price standard deviation, but also a parallel decrease in trading activity. 
Such evolution drove the slump of fixed-income revenues and 
the decline in profits of large dealers, as already underlined by
\citet{BMV}.

In Europe, the observed evolution in the U.S.A. bond market  after the introduction of TRACE  
caused a lively debate within European institutions 
\citep{Glover},
with consequent delay in the enforcement of mandatory transparency rules in 
the European Union.\footnote{European companies had the equivalent of about \euro 8.4 trillion of bonds outstanding in various currencies in May 2014, 
up from \euro 6.3 trillion at the beginning of 2008, so that the European bond market is almost as large as that of the U.S.A. \citep{Glover}. }
After several years of haggling between policy makers,
the ruling of bond market transparency was included within the update of the Markets in Financial Instruments Directive (also known as “MiFID II”) 
approved in April 2014 and binding since January 2018. 
The European Securities and Markets Authority (ESMA) is in charge of collecting transaction data from dealers and disclose information on bond liquidity. 
Since compulsory data collection started only in January $2018$, it is too early to draw significant conclusions from the analysis of time series of ESMA transaction data. 

\bigskip

The econometric analyses  
show evidence of a split market, with large differences in the liquidity of debt securities traded in the same marketplace.
Moreover, the bond market can be very differentiated even 
for the same issuing institution: 
some bonds can be very illiquid  while some others, even with similar characteristics (e.g. the same time to maturity), trade several times every day, 
with trading activity far from being uniform over time but mostly concentrated on recently issued bonds (`on-the-run' issues).
These stylized facts clarify the relevance of
the {\it sheer liquidity} premium investigated by \citet{Helwege};
in a framework where the effectiveness of econometric techniques is hampered by the sparsity of trades 
and by non-stationary data, 
we deem useful to resort to a theoretical model that distinguishes 
between the credit and the liquidity components of the spread. 
This leads us to a first 
research question: 
could
the estimation 
of the {\it sheer liquidity} spread in bond yields  allow 
to pick out the relevant risk factors
that can be measured in observable market data (e.g. the volatility of the bond)?
The identification of the risk factors relevant in the liquidity spread could point to
risk measures (i.e. the corresponding sensitivities) that 
allow to monitor and to control the risks in an illiquid corporate bond portfolio. 

Moreover, the above mentioned econometric studies considered only a small fraction of TRACE bond transactions; 
they focus on the bonds that trade more regularly: for example, 
\citet{NFL}
limited the analysis to the 20.6\% of the total number of bonds in TRACE dataset,
\citet{Schestag2016}
selected the bonds that trade at least 75\% of trading days during their life span (i.e. the 5.5\% of the total), 
\citet{Helwege} 
considered only the bonds that trade at least four times a day, selecting 4.2\% bonds within the total TRACE set.
A second relevant 
research question arises: can we provide a price to the most illiquid bonds, i.e. the ones neglected in
econometric analyses? 
A theoretical approach could focus on these illiquid bonds and 
suggest a price
when liquid bonds for the same issuer are present in the market.
This question could be relevant for practitioners, not only when pricing corporate bonds 
but also when setting haircuts for illiquid bonds accepted as collateral.

Finally, as already pointed out in existing studies, 
sparse data in the corporate bond market 
often prohibit the use of liquidity 
metrics and 
approaches designed for 
the equity market.\footnote{Corporate bonds present 
differences of some orders of magnitude w.r.t. large cap stocks:
``a typical US large cap stock, say Apple as of November 2007, had a daily turnover of around 8bn USD'' with an
``average of $6$ transactions per second and on the order of 100 events per second affecting the order book'' \citep[cf.][p.76]{BFL}.}
These sparse data should imply 
a change of paradigm that privileges parsimony and simplicity.
For this reason, 
on one side, parsimony suggests to consider a reduced-form model \citep[see, e.g.][]{DuffieSingleton1999, Schonbucher1998} that allows direct calibration of model parameters, 
and, on the other side, 
simplicity leads back to the question  we started with in this Introduction, i.e. on the opportunity to 
address just one single aspect of market liquidity: 
the time-to-liquidate a given position (hereinafter ‘ttl’).
Our theoretical approach presents several advantages: liquidity is considered an intrinsic characteristic of each single issue, 
it can vary over time and it depends on the size. 
Liquidity is expressed in terms of a price discount (or equivalently in terms of a liquidity spread) as a simple closed formula 
that depends on a single 
parameter, the ttl.
This is the time lag that, at a given date and for a given size, an experienced bond trader needs to liquidate the position. 

\bigskip
Theoretical studies on bond liquidity are rather few.
Our approach is reminiscent of the celebrated work of 
\citet{Longstaff}
on the non-marketability.
The brilliant idea of Longstaff has been to view liquidity as a right (and then as an option) in the hands of the asset holder: 
if an asset is liquid the holder can sell it at any time in the market. 
Therefore, liquidity can be priced as a derivative. In particular, Longstaff considered
non-dividend-paying shares in IPOs  in an equity market. 
Following the option idea of Longstaff, \citet{KoziolSauerbier}
tackle with a numerical technique a liquidity problem in the case of a risk-free zero-coupon (ZC) bond 
with a simple model \citep{Vasicek} that includes interest rate dynamics but neglects credit risk.
Two are the theoretical approaches that include also credit risk, both considering a structural-model for corporate bonds:
\citet{EricssonRenault} capture both liquidity and credit risks through exogenous liquidity shocks;
\citet{TychonVannetelbosch} describe liquidity endogenously modeling
investors with  heterogeneous valuations about bankruptcy costs. 
These two papers showed --in a Nash bargaining setup--
how bond prices are influenced by both  liquidity and renegotiation/recovery in financial distress.
Unfortunately, their parameter rich structural-model 
allows  only a numerical solution and it is not simple to be calibrated on real data, 
because it includes some unobservable parameters.  

In this paper we propose a 
reduced-form modeling approach --for the first time in a study on corporate bond liquidity-- that, 
on the one hand, 
allows an analytical solution for the liquidity spread in presence of default risk and coupon payments,
and on the other hand,
reproduces the simple calibration features of reduced-form models.

\bigskip
We consider an application to the European market, where the problem of pricing illiquidity is even more significant than in the American one, 
due to the differences in mandatory transparency requirements 
mentioned above. 
Moreover, in Europe, it is relatively frequent to observe private placements to institutional investors, 
where a single issue is detained by a very limited pool of bondholders, and, especially in the financial sector, there are several bonds with small issue sizes aimed either at retail investors or at 
private-banking clients of a banking institution. Often no market price is available for several days
and a closed formula can be relevant and useful in these cases. 
The proposed formula, besides bond characteristics (maturity, coupon, sinking features, etc...), depends on standard market quantities, such as i) the observed risk-free interest curve ii) issuer's credit spread term-structure and iii) bond volatility. 
In particular, via a detailed calibration on interest rate and credit market data
for two European issuers in the financial sector,
we show
the relative importance of model parameters, such as volatility and time-to-maturity, 
in liquidity spreads. 

\bigskip

The contributions of this paper to the existing literature on illiquid corporate coupon bonds are threefold. 
First, 
 it identifies the {\it sheer liquidity} premium as the key quantity that relates 
 liquid and illiquid issues of the same corporate issuer.
It clarifies the role played in the liquidity spread 
by the time-to-liquidate a position and by the bond volatility. 
Second, 
the elementary  model set-up allows to
deduce a closed formula for illiquid corporate coupon bonds.
It provides  illiquid bond prices when they are not available in the marketplace.
Third, this paper introduces realistic corporate bond features as 
coupon payments and defaults via a reduced-form model.
The proposed parsimonious modeling approach allows a calibration on market data:
we show two examples in the European market.

\bigskip

The remainder of this paper is organized as follows. 
In Section 2, we describe the model set-up
and the liquidity problem formulation.
In Section 3, we deduce the closed formula and in Section 4, we show how to calibrate the model parameters on 
 real market data for two European bond issuers. In Section 5, we make some concluding remarks.

\section{The model}

The model includes two sets of financial ingredients:
on one side, the model set-up for the interest rate and the credit components in
liquid corporate bonds, and 
on the other side, a description on how illiquidity affects corporate bond prices.

This section is divided into three parts.
In the first subsection we recall the modeling framework for 
corporate bonds \citep{DuffieSingleton1999, Schonbucher1998}, 
while in the following we introduce illiquidity.
In the last subsection we specify a parsimonious dynamics for  interest rates and credit spreads, 
that will allow to 
obtain the closed formula for illiquid corporate bonds in Section  3.

\subsection{The modeling framework for liquid bonds}

We model interest rates and credits according to a zero-recovery model introduced by
\citet{DuffieSingleton1999} and \citet{Schonbucher1998}.
This reduced-‐form  model is a generalization of the model of \citet{HJM} to the defaultable case 
and it was named by \citet{DuffieSingleton1999} the Defaultable HJM framework
(hereinafter, DHJM).\footnote{The model is also known as Duffie-Singleton model.}
As we underline in this subsection, the DHJM is a flexible modeling framework depending on the chosen volatility structure:
in subsection 2.3 we select a particular model within this framework that allows a simple closed formula for illiquid corporate bonds 
and an elementary calibration.

\bigskip

In this subsection we briefly recall the DHJM,
we describe the dynamics for a corporate bond and a 
forward contract written on it. 
We use a notation very close to  \citet{Schonbucher1998}
that is similar to the one  in standard textbooks 
\citep[see, e.g.][Ch.5 and 6]{Schonbucher}. 

\bigskip

The DHJM is a standard intensity based model, 
where  the default for a corporate obligor $C$ is modeled via a jump of a process ${\cal N}_t$ with intensity $\lambda_t$ \citep[see, e.g.][]{Schonbucher}.


Market practitioners view corporate bond spreads via Zeta-spreads: 
from a modeling perspective, this corresponds
to  considering zero recovery and to stating that the default probability models the whole credit risk for the obligor $C$. 
In particular we consider a zero-recovery model as a limit case of a fractional-recovery model.
A (liquid) defaultable ZC with Fractional Recovery (FR) between time $t$ and maturity $T$, $\overline{B}_{q}(t, T)$, is the price of a defaultable ZC where, 
if a jump
occurs at $t$,  
the value of the defaultable asset is $1-q$ times its 
pre-jump
value, with $0<q<1$, i.e.
\be
\label{eq:FR main}
\overline{B}_{q}(t, T) =  (1-q) \overline{B}_{q}(t^-, T) \; .
\en
A defaultable ZC with zero-recovery, $\overline{B}(t, T)$, can be seen as a particular case of a ZC with FR when $q$ tends to $1$ 
\citep[see, e.g.][Ch.6]{Schonbucher}. 
Often, it is simpler to use this modeling perspective for a generic $q$ and then consider the case with $q$ close to $1$. 
This is the approach we follow in this paper.

\smallskip

The only difference of DHJM w.r.t. standard intensity based (or reduced-form) models
is that intensity as well as interest rates 
are not determinstic functions 
but follow continuous stochastic dynamics, driven in general by 
a $d$-dimensional vector of correlated Brownian motions ${W}_t$, i.e. one has $d W^{(j)}_t \, d W^{(l)}_t = \rho_{j l} \, dt$  for $j, l = 1 \ldots  d$ 
and $\rho \in \Re^{d \times d}$ the instantaneous correlation matrix.  
In probability theory such a process is called a Cox process. 
The great advantage of this modeling framework is that, 
on one side, it models defaults in an elementary way as standard reduced form models, 
and, on the other side,
it does not impose 
to have deterministic interest rates and intensities but it allows to model both of them via continuous (stochastic) dynamics.

\smallskip

A little of terminology can be useful.
The information relative to the continuous paths of interest rates and intensities up to time $t$ 
(and then the knowledge of the  Brownian motions in their dynamics up to $t$) is indicated with $ {\cal G}_{t}$,
while the whole  information --i.e. including even the jumps that have occurred-- up to time $t$ is indicated with $ {\cal F}_{t}$.
Note that this terminology is useful when we have expectations,
that can be conditioned on either $ {\cal F}_{t}$ or  $ {\cal G}_{t}$:
in practice this terminology is useful because it allows to generalize  well known properties of financial quantities that evolve continuously in time \citep[see, e.g.][]{Musiela}
to defaultable quantities as corporate bonds.

\bigskip

Two are the main properties of DHJM:  it allows 
i) to relate interest rates and bond prices in both risk-free and defaultable settings and 
ii) to indicate the dynamics of ZCs.

First, risk-free ZC, ${B}(t_0, T)$, and the risk-free rate $r_t$
are related via
the stochastic discount 
${D}(t_0, T) := \exp - \left( \int^T_{t_0} r_s ds \right)$ 
\[
{B}(t_0, T) := \mathbb{E} \left[{D} (t_0, T) | {\cal F}_{0} \right] \;\; .
\]
Defaultable quantities with fractional recovery $q$ are introduced in a similar way. 
The defaultable rate $\overline{r}_t := r_t + q \lambda_t$ and
the
defaultable ZC $\overline{B}_q(t_0, T)$   are related via
\[
\overline{B}_q (t_0, T) := \mathbb{E} \left[{D} (t_0, T) (1-q)^{{\cal N}_T} | {\cal F}_{0} \right] = 
\mathbb{E} \left[\overline{D}_q (t_0, T) | {\cal G}_{0} \right]  \; ,
\]
where the defaultable  stochastic discount is
$\overline{D}_q(t_0, T) := \exp - \left( \int^T_{t_0} \overline{r}_s ds \right) $.

\smallskip

Second, in the DHJM, the dynamics for ZCs under the risk-neutral measure are for a generic $t \in (0,T]$
\be
\left\{
\begin{array}{lcl}
{\displaystyle \frac{d B(t, T)}{B(t, T)}} & := &  r_t  \, dt + {\sigma} (t,T) \cdot d {W}_t  \\[4mm]
{\displaystyle \frac{d \overline{B}_{q}(t, T)}{\overline{B}_{q}(t^-, T)}} & := & \overline{r}_t dt + \overline{{\sigma}} (t,T) \cdot d {W}_t - q \, d {\cal N}_t \; \;  \\
\end{array}
\right.
\label{eq:DHJM}
\en
with $B(t_0, T)$ and $ \overline{B}_q(t_0, T)$ 
their initial conditions at value date $t_0$ \citep[cf. e.g.][p.173, eqs. (46) and (44) in the zero-recovery case]{Schonbucher1998}.
The volatilities $ \sigma(t,T) $ and $  \overline{\sigma}(t,T)$ are $d$-dimensional
vectors
with $ \sigma(T,T) = \overline{\sigma}(T,T) = {\mathbf 0} \in \Re^d$.
We indicate 
with $ x \cdot y $  the scalar product between two vectors $x, y \in \Re^d$ and
with  $ x^2 $  the scalar product  $ x \cdot \rho \, x$, 
$x \in \Re^d$ and $\rho \in \Re^{d \times d}$ the instantaneous correlation introduced above.  
As already mentioned, 
the rates $r_t$ and $\overline{r}_t$ are described by continuous stochastic differential equations: 
their dynamics is reported in Appendix {A} together with some basic relations that hold for DHJM.


\bigskip

Let us introduce a simple derivative contract that will play a key role when modeling illiquidity.

The forward defaultable ZC bond at time $t$ is a derivative contract with a reference obligor $C$ and characterized by three times $t$, $\tau$ and $T$ 
s.t. $t\le \tau \le T$. This forward contract is characterized by 
the payment in $\tau$ of an amount  in order to receive in $\tau$ a ZC with maturity $T$. 
This amount is equal to a fraction of a price $\overline{B}(t; \tau, T) $ established in $t$, 
where the fraction depends on the number of jumps that occur up to time 
$\tau$.\footnote{This fraction is equal to $(1-q)^{{\cal N}_{\tau}}$, the same price reduction incurred  by the corresponding corporate bond  up to $\tau$.} 
%
This  price $\overline{B}(t; \tau, T)$  is related to a defaultable ZC via
\be
\overline{B}(t; \tau, T) = \frac{ \overline{B}_q(t, T) } { \overline{B}_q(t, \tau) } \;\; .
\label{eq:forwardZC}
\en
This is the unique price that does not allow arbitrage in the DHJM, as it can be shown via direct computation.
Moreover, the forward defaultable ZC presents the property that  $\overline{B}(t; t, T) = \overline{B}_q(t, T) $, i.e. the forward defaultable bond price tends
to the defaultable bond price as time $\tau$ tends to $t$.
This contract presents interesting financial features, as we will discuss in the zero-recovery case, and a simple dynamics in the DHJM

\be
\displaystyle \frac{d \overline{B}(t; \tau, T)}{ \overline{B}(t^-; \tau, T)} = \displaystyle \frac{d \overline{B}(t; \tau, T)}{ \overline{B}(t; \tau, T)} =
 \left[  \overline{{\sigma}}(t,T)  -  \overline{{\sigma}}(t,\tau)\right] \cdot  \left[ d {W}_t  +  \rho \, \overline{{\sigma}}(t,\tau)  \, dt \right] \; \;  t \in [0, \tau] \;
\label{eq:ForwardDinamicsI}
\en

This property can be deduced using 
the Generalized It\^{o} Lemma \citep[see, e.g.][Ch.4, p.100]{Schonbucher}, the dynamics  (\ref{eq:DHJM})  and equation (\ref{eq:forwardZC}).

\bigskip

Equation (\ref{eq:ForwardDinamicsI}) states that the dynamics of the forward  defaultable ZC bond price  is continuous and does not depend on the fraction $q$:
it is, {\it mutatis mutandis}, 
the same 
as the corresponding dynamics for a risk-free forward ZC bond \citep[see, e.g.][]{Musiela}.

\bigskip

It is possible to introduce a $\tau$-defaultable-forward measure (hereinafter also $\overline{\tau}$-forward measure), s.t.
the process
\[
 {W}^{(\overline{\tau})}_t :=  {W}_t  + \int^t_{t_0} \rho \, \overline{{\sigma}}(s,\tau)  \, ds
\]
is a $d$-dimensional Brownian motion under the new measure. 
We indicate with $ \mathbb{E}^{(\overline{\tau})} \left[ \; \bullet \; \right]$ the expectation under the 
$\overline{\tau}$-forward measure.
A consequence of equation (\ref{eq:ForwardDinamicsI}) is that,
in the $\overline{\tau}$-forward measure, 
the dynamics for the forward defaultable ZC $\overline{B}(t; \tau, T)$ has a particularly simple form:
\be
\displaystyle d \overline{B}(t; \tau, T) = \overline{B}(t; \tau, T) \,
 v(t; \tau, T)  \cdot   d {W}^{(\overline{\tau})}_t  
\label{eq:ZCdynamycs}
\en
with $v(t; \tau, T) := \overline{{\sigma}}(t,T)  -  \overline{{\sigma}}(t,\tau) $.

\bigskip

Hereinafter, we consider the zero-recovery model, obtained as a limit case of the fractional-recovery model for  $q=1^-$.
The zero-recovery model allows to simplify the notation.
Defaultable quantities with zero-recovery are indicated as the defaultable quantities with fractional recovery
 without the subscript $q$, i.e. 
$\overline{B}(t, T) := \overline{B}_{q=1^-}(t, T) $ and 
$\overline{D}(t, T) := \overline{D}_{q=1^-}(t, T)$. 
Their relation at value date becomes $\overline{B} (t_0, T) = \mathbb{E} \left[{D} (t_0, T) \mathbbm{1}_{t_d > T} | {\cal F}_{0} \right]$,
with $t_d$ the default time, that corresponds to the first jump of $\{{\cal N}_t \}_{t\ge0}$.

\begin{figure}
  	\begin{center}
      	\includegraphics[width=0.70\textwidth]{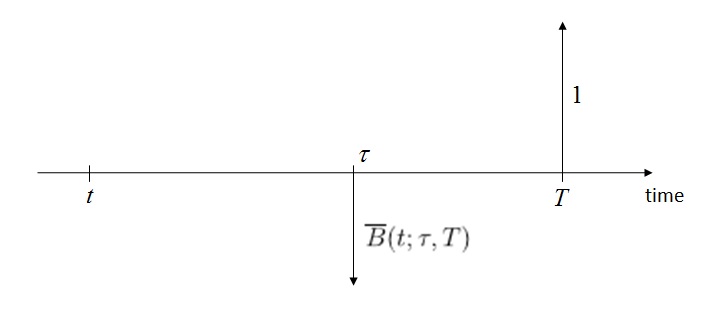}
	\end{center}
  	\vspace{-15pt}
  	\caption{\small We show the flows that characterize, in the zero-recovery case, a long position on a  forward defaultable ZC bond paid at $\tau$ if no default 
event occurs up to $\tau$. A forward defaultable ZC 
 price is established at time $t$. The contract gives the right to receive $1$ if no default event occurs up to $T$.}
\label{fig:ForwardBond}
\end{figure}

Moreover, in the zero-recovery case  
the forward defaultable ZC bond $\overline{B}(t; \tau, T)$ becomes an elementary contract.
It is characterized by the payment in $\tau$ 
of
\[
\left\{
\begin{array}{ll}
\overline{B}(t; \tau, T) & \text {if the obligor } C \text{ has not defaulted up to time } \tau  \text{ and} \\
 0 & \text{otherwise} 
\end{array}
\right.
\] 
 in order to receive $1$ in $T$ if the obligor $C$ has not defaulted up to time $T$ 
(and zero otherwise), where the price  $\overline{B}(t; \tau, T)$  is established in $t$.\footnote{
As underlined in \citet[p.165]{Schonbucher1998} this derivative is not a ``classical" $\tau$-forward contract
and it can be replicated as a portfolio of defaultable ZC bonds.} 
In Figure \ref{fig:ForwardBond} we show the flows that characterize a  forward defaultable ZC bond in the zero-recovery case.

\bigskip

In this study we focus on fixed rate coupon bonds that are not callable, puttable, or convertible. 
A (liquid) corporate coupon bond of the obligor $C$ 
is
\be
\overline{P}(t_0; {\bf c}, {\bf t}) := \sum^{N}_{i=1} c_i \overline{B} (t_0, t_i) \; .
\label{eq:CouponBond}
\en 
In the definition of a corporate coupon bond (\ref{eq:CouponBond}), the price depends on the set of flows $ {\bf c} := \{ c_i \}_{i=1\ldots N}$ 
and the set of payment dates $ {\bf t} := \{ t_i \}_{i=1\ldots N}$. 
We indicate with $T$ the bond maturity, i.e. $t_N = T$.
The $i^{th}$ payment $c_i$ at time $ t_i$ for ${i<N}$ is the coupon payment with the corresponding daycount, 
while the last payment at $t_N$ has the bond face value added to the coupon payment.
A corporate coupon bond $\overline{P}$ 
always indicates invoice (or dirty) prices, as in standard fixed income modeling.

We indicate with $\overline{P}(t, \tau; {\bf c}, {\bf t})$ the forward defaultable coupon bond in $t$ paid in $\tau$ 
that generalizes the forward defaultable ZC (\ref{eq:forwardZC}) to the case with coupons (\ref{eq:CouponBond})
\be
\overline{P}(t, \tau; {\bf c}, {\bf t}) =  \sum^{N}_{i=1; t_i > \tau} c_i \frac{\overline{B} (t, t_i)}{\overline{B} (t, \tau)} \;\; ;
\label{eq:ForwardCouponBond}
\en
where
in the forward  $\overline{P}(t, \tau; {\bf c}, {\bf t})$ only coupons with payment dates $t_i > \tau $ appear.

\bigskip

In the next subsection we describe 
how illiquidity affects corporate coupon bonds (\ref{eq:CouponBond}). 

\subsection{The sheer liquidity premium}

This subsection focuses on the main modeling assumption and it is the core of the approach we propose for pricing illiquid corporate bonds.
As already stated in the Introduction, 
we model illiquidity following closely the option approach of \citet{Longstaff}.
We consider a hypothetical investor who holds, at value date $t_0 =0$, 
an  illiquid corporate bond. 
The illiquidity is characterized by one main property:
the investor needs some time  in order 
to liquidate a position with a given size of that bond. 
This hypothetical investor will be able to sell the position in
the illiquid bond
only after a time-to-liquidate $\tau$  at the same price as 
a liquid bond with the same characteristics 
(issuer, coupons, payment dates).

We assume that this investor is an {\it experienced trader} better informed than other market players on that particular corporate market segment: 
this {\it experienced trader} knows all the features of the bonds of that issuer and all the potential clients that could be interested in buying the bond he holds.

After the seminal paper of \citet{Kyle1985}, 
the assumption that some market players are better informed than others
is rather common when analyzing, 
from a theoretical perspective, specific trading mechanisms and the price formation process of some assets.
In particular, Longstaff's idea is simple and brilliant: this {\it experienced trader} 
``has {\it perfect market timing} ability that would allow him to sell the security and 
reinvest the proceeds in the riskless asset, 
at the time $t$ that maximizes the value of his portfolio. 
[...] As long as the investor cannot sell the [illiquid]
security prior to time $\tau$, however, he cannot benefit from having perfect
market timing ability. (...)
[Illiquidity] 
imposes an important opportunity cost on this
hypothetical investor" 
\citep[cf.][pp.1768-1769]{Longstaff}. 

Summarizing, ``this incremental cash flow can also be viewed as the payoff from an option" \citep[cf.][p.1769]{Longstaff}.
Liquidity is seen as an incremental right (i.e. an option) in the hands of the investor, 
who can
liquidate a given position at market price whenever he desires. 
The additional value of the liquid security over the illiquid one is calculated by regarding the optimal strategy of
this hypothetical investor.

\bigskip

More in detail, two are the differences/specifications of our approach w.r.t. \citet{Longstaff}, 
due to the fact that we focus on an illiquid defaultable corporate bond. 
 
First, as an example, \citet{Longstaff} focuses his attention on a non-dividend paying stock in IPOs: 
the value of the asset sold at time $t\in(t_0, \tau)$ 
including the reinvestment up to $\tau$  is nothing else than a forward contract  in $t$ and expiry in $\tau$ on the asset.
In this paper, we follow this approach, dealing with a derivative rather than the underlying asset:
the contract we consider is the forward defaultable coupon bond (\ref{eq:ForwardCouponBond}) introduced in previous subsection.  

Second, we require only an {\it almost perfect market timing} for the hypothetical investor. 
The {\it experienced trader} has two relevant pieces of information:
\begin{enumerate}
\item he knows whether the corporate will default before $\tau$ (but he does not know exactly when) and 
\item in case of no default up to the time-to-liquidate $\tau$, he has perfect market timing ability on the forward defaultable coupon bond
before  $\tau$.
\end{enumerate}
We return to 
the plausibility of these hypotheses in the following.


\bigskip

Hence, when dealing with fixed income securities, we have to consider that bonds pay coupons and that they can default.
Two are the cases of interest: either the corporate issuer defaults before the time-to-liquidate $\tau$ or it defaults after $\tau$. 

In the former case, 
the illiquid position 
has a value equal to zero in $\tau$ (and then also in the value date), 
while the liquid bond is sold immediately at its price at value date $\overline{P}(t_0; {\bf c}, {\bf t})$.
This is a consequence of the fact that the hypothetical investor knows that the corporate will default before $\tau$ but he does not know exactly when;
he will liquidate his liquid position as soon as possible, while the illiquid one returns zero, due to the zero-recovery of the bond. 

In the latter case, 
the {\it experienced trader} sells the 
liquid position via the forward defaultable coupon bond. It sells this forward with optimal timing, i.e. the selling price is
\[
M_\tau := \max_{t_0 \le t \le \tau} \overline{P}(t, \tau; {\bf c}, {\bf t}) \; \; ;
\]
this price is received at time $\tau$ (because it is a forward defaultable bond, see also Figure \ref{fig:ForwardBond}).
He also sells the illiquid bond in $\tau$ at $ \overline{P}(\tau; {\bf c}, {\bf t})$.
Summing up, the two possibilities for the liquid and illiquid bonds are shown in Table \ref{tab: Possibilities}.

\begin{table}[h!]
\begin{center}
\[
\begin{array}{l|cc|l}
& liquid & illiquid &   \\[1mm] \hline  
& & & \\[-2mm]
t_d \le \tau & \overline{P}(t_0; {\bf c}, {\bf t}) & 0 & \text{received in } t_0 \\[2mm]
t_d > \tau & M_\tau & \overline{P}(\tau; {\bf c}, {\bf t})  & \text{received in } \tau \\[2mm] \hline
\end{array}
\]
\end{center}
\caption{\small We show the value of liquid and illiquid bonds for an {\it experienced trader}. 
On the one hand, in case of default before $\tau$ (i.e. $t_d \le \tau$ ) the illiquid bond has no value --due to the zero recovery-- while the liquid one is sold immediately
at its price in $t_0$. On the other hand, in case of default after $\tau$ (i.e. $t_d > \tau$ ),
the illiquid bond is sold at $ \overline{P}(\tau; {\bf c}, {\bf t})$ and the liquid position via the forward defaultable coupon bond (with optimal timing): 
both prices are received in $\tau$.
}
\label{tab: Possibilities}	
\end{table} 

Longstaff's idea is very intuitive: the main limitation of holding an illiquid bond, compared with a comparable issue of the same corporate entity, is related to the impossibility for a while to sell the bond and convert its value into cash.
The time-to-liquidate $\tau$ is the main exogenous model parameter: it models the liquidity restriction as an opportunity cost for this hypothetical investor.
We can now state the main assumption of our modeling set up.

\bigskip

{\bf Assumption}: 

\smallskip

The {\it sheer liquidity} premium $ \Delta_\tau $ is defined as the value at $t_0$ of the difference between liquid and illiquid prices
of  bonds of the same corporate issuer with the same characteristics (same coupons and payments dates).  
Its present value equals 
\be
 \Delta_\tau :=
\mathbb{E}
\left[ 
{D} (t_0, \tau) \mathbbm{1}_{t_d > \tau}  
\left( M_\tau  - \overline{P}(\tau; {\bf c}, {\bf t})  \right)  +  \mathbbm{1}_{t_d \le \tau}  \overline{P}(t_0; {\bf c}, {\bf t}) | {\cal F}_{0}  
\right] \; .
\label{eq:model}
\en

The {\it  sheer liquidity} premium is equal to the sum of two terms. 
As shown in Table \ref{tab: Possibilities},
the first term, in case of no default up to  the time-to-liquidate $\tau$, is equal to the difference, 
for the hypothetical investor, of the selling prices in $\tau$ of the liquid and illiquid forward defaultable bonds; 
the second one, 
in the event of default before $\tau$,
is  equal to the liquid defaultable bond price in $t_0$. $ \qquad\hspace*{\fill} \diamondsuit \; $

\bigskip

Let us discuss the plausibility of the above assumption and the role of the {\it experienced trader}.

The {\it experienced trader} closely models some real market makers that operate
in the corporate bond market. 
Within this market some market makers, often within major dealers,
are very specialized on few issuers and sometimes only on few issues of 
a given issuer.
On the corporate side, 
these traders often know personally the top-management of the firm,
its liquidity needs, its funding policy, 
the next issues in the primary market. 
On the market side,
they know in detail the market segment where they operate, 
every relevant investor who has invested on that corporate or who can be interested to that firm name,
they advise the company to obtain financing in the primary market
overseeing primary bond sales to investors 
in that company or in strictly related firms.\footnote{Market making for corporate bonds is very concentrated: for each bond often there are few prominent dealers, 
one often being bond underwriting dealer. This is in line with \citet{OWZ}, who find --in the U.S.A. corporate bond market-- that although 
there are several hundred dealer firms, in many bonds there are only one to two active dealers per year, with the top dealer doing 
on average 69\% of the volume and the top two dealers having a 85\% market share for the average sample bond.}

As previously mentioned, two are the main hypotheses on the {\it experienced trader} related to
i) the knowledge of the arrival of a default before $\tau$ and 
ii) the perfect timing ability in case of no default before $\tau$.
Both hypotheses sound reasonable for market makers who have access to the soft information we have  described above.

\smallskip

Moreover, something can be said on the first  hypothesis;
it can be useful to remember that the time-to-liquidity is few months in the most illiquid cases and
some days or weeks more generally, and 
most of the 
trading activity on illiquid corporate bonds is concentrated on 
issuers either investment grade or with
the highest ratings in the speculative grade.
Within such a short time interval (compared to the typical maturity of corporate issues), 
it sounds reasonable --for this class of market makers-- 
to know  with high-probability whether the corporate issuer would default or not 
in the (rare) event of issuer default before the time-to-liquidate; it is instead rather improbable that they know exactly when.

\smallskip

The second hypothesis is the same as Longstaff's one, under the condition of no-default before $\tau$. 

For sure,
in the corporate bond market, asymmetric information helps 
these market makers in the selection of market timing, where   
higher prices are mainly due to
a renewed interest on a specific firm related to, e.g., 
a change in its financing policy, new issues in the primary market  or
unexpected reported results by the corporate.
It is also sure that this information sounds more valuable in a market, as the 
corporate bond one, 
where market abuse is extremely difficoult to detect.

\bigskip

Having said that, though the {\it experienced trader} is an idealization.
As already stated by Longstaff, the {\it sheer liquidity} premium $\Delta_\tau$
``would be less for an actual investor with
imperfect market timing ability. Thus, the present value of the incremental
cash flow represents an upper bound on the value of marketability"\citep[cf.][p.1769]{Longstaff}.

\bigskip   

{\bf Remark.}
The above definition does not consider the case when coupon payments take place between the value date $t_0$ and $\tau$. 
We have already underlined that the time-to-liquidate
is, even in the most illiquid cases, a few months, and then at most one coupon payment could be present in the time interval  $(t_0, \tau)$.
The first coupon, when paid before $\tau$,  can be separated by the other flows in the coupon bond; 
a technique known in the market place as coupon stripping.
In practice, corporate bond traders consider that payment, i.e. within a short lag in the future, very liquid. 
We assume that this coupon makes 
the same contribution to both the liquid and illiquid coupon bonds, i.e.
it maintains only its interest rate and credit risk components; thus, this coupon does not appear in the {\it sheer liquidity} premium $ \Delta_\tau $ in (\ref{eq:model}). 
Hereinafter, we consider in the corporate coupon bond only the coupons after the time-to-liquidate, 
i.e. in definition (\ref{eq:CouponBond}) the first coupon in the sum is the first one paid after  $\tau$.

\subsection{A parsimonious model selection}

It can be interesting to observe that,
within the DHJM framework, the {\it sheer liquidity} premium (\ref{eq:model}) can be written in a simpler form 
\be
\label{eq: lemma3}
\begin{split}
\Delta_\tau 
& = \overline{B} (t_0, \tau) \,
\mathbb{E}^{(\overline{\tau})} \left[ M_\tau | {\cal G}_{0} \right] -
\mathbb{P}(t_0, \tau) \, \overline{P}(t_0; {\bf c}, {\bf t}) \;\; ,
\end{split}
\en 
where $\mathbb{P}(t_0, \tau)$ is the issuer survival probability up to the time-to-liquidate 
(for a deduction of this simple equality within DHJM, see also (\ref{eq: lemma3Appendix}) in Appendix {A}).
Let us notice that the first term of (\ref{eq: lemma3}) 
is the only quantity in the price of illiquidity $\Delta_\tau$ 
rather complicated to be computed: 
it depends on
$\overline{r}_t$ and not separately on 
$r_t$ and $ \lambda_t $.

\bigskip

This property holds whichever zero-recovery DHJM model is selected 
(i.e. whatever $\sigma (t,T) $ and $\overline{{\sigma}} (t,T)$  are chosen)
for the dynamics (\ref{eq:DHJM}) of 
$ B(t, T) $ and 
$ \overline{B}(t, T) $. 
As discussed in the Introduction, the main driver for model selection is parsimony 
when dealing with illiquid corporate bonds, due to the poorness of the data set and model calibration issues. 
One of the simplest and most parsimonious models within DHJM 
was proposed by \citet{SchonbucherTree}, where both $r_t$ and $\lambda_t $ follow two correlated 1-dimensional \citet{HW1990} models 
\[
\left\{
\begin{array}{lcl}
r_t & = & \varphi_t + x^{(1)}_t \\
\lambda_t & = & \psi_t  + x^{(2)}_t 
\end{array}
\right.
\]
where $ x^{(1)}_t $ and $ x^{(2)}_t $ are two correlated Ornstein--Uhlenbeck processes (OU) with zero mean and zero initial value;
$\varphi_t$ and $\psi_t$ are two deterministic functions of time. 
This model has the main advantage of allowing an elementary separate calibration of the zero-rates (via $\varphi_t$) and the Zeta-spread  curve (via $\psi_t$) 
at value date $t_0$.
A consequence of the observation that only the dynamics for $ \overline{r}_t $ matters for liquidity,
makes us consider an even simpler model with the two OU perfectly correlated, i.e. with  only
one OU driver, as recently proposed by \citet{BavieraSwaption} in a multi-curve problem.

\bigskip


In this case,
 the risk-free interest rate $r_t$ and the intensity $\lambda_t$ are modeled as
\be
\left\{
\begin{array}{lcl}
r_t & = & \varphi_t + (1- {\hat \gamma}) \; x_t \\
\lambda_t & = & \psi_t  + {\hat \gamma} \; x_t 
\end{array}
\right.
\label{eq: dynamics} 
\en
with  
$x_t$ an OU with zero mean and initial value
\[
\left\{
\begin{array}{lcl}
 d x_t & = & - {\hat a} \, x_t \, dt + {\hat \sigma} \, d W_t \\
 x_{0} & = & 0
\end{array}
\right.
\]
where ${\hat \gamma} \in[0,1]$, while $\hat{a}, \hat{\sigma}$ 
are two positive constant parameters.

\bigskip

This model selection is in line with day-to-day practice.
In the marketplace, often one cannot observe 
options
that allow calibrating separately the volatility 
of the risk-free curve and the volatility of the credit spread,
i.e. there is not enough information to discriminate between the two dynamics.
One can associate a fraction ${\hat \gamma}$ of the total dynamics to the credit component and the remaining fraction to the interest rate component.
Conversely, the two initial curves (risk-free zero-rate and Zeta-spread) can be easily calibrated separately on market data and the integrals of
$ \varphi_t, \psi_t $ between $t_0$ and a given maturity $T$ are related to these two curves up to $T$. 
We provide final formulas in terms of $ B(t_0,T)$ and $ \overline{B}(t_0,T)$ because
both curves can be calibrated directly from market data.

\bigskip

A consequence of (\ref{eq: dynamics}) is that the defaultable rate $ \overline{r}_t = r_t + \lambda_t$ is
\[
 \overline{r}_t = \varphi_t + \psi_t + x_t \; \; .
\]
It is modeled according to a  Hull--White 
model with one-dimensional volatility \citep[see, e.g.][]{BM}
\be
 \overline{\sigma}(t,T) = \frac{\hat \sigma}{\hat a} \left( 1 - e^{- {\hat a}(T-t) }\right) \, \in \, \Re \qquad t \le T \, 
\label{eq:HW volatility}
\en
and then the volatility $ v(t; \tau, t_i) = \overline{{\sigma}}(t,t_i)  -  \overline{{\sigma}}(t,\tau) $, 
defined in equation (\ref{eq:ZCdynamycs}), is a separable function in the times $t$ and $t_i$, i.e. 
\be 
v(t; \tau, t_i) = \zeta_i \, \nu(t) 
\label{eq:Volatility}
\en 
with 
$\zeta_i := (\hat{\sigma}/\hat{a}) \left[ 1 - e^{- \hat{a} (t_i - \tau) } \right]$ and 
$\nu(t) := e^{- \hat{a} (\tau -t)}$ with $ t_0 \le t \le \tau \le t_i$.  

\bigskip

In the next section we show that, 
with the proposed model (\ref{eq: dynamics}),
it is possible to compute the {\it sheer liquidity} premium $\Delta_\tau$ (\ref{eq: lemma3})  via a closed formula and
it is possible to associate a liquidity spread as a component of the corporate bond spread in addition to the credit spread.


\section{A closed formula for illiquid corporate coupon bonds}

In this section we present the main result of this paper: 
the {\it sheer liquidity} premium $\Delta_\tau$ 
(\ref{eq: lemma3}) can be evaluated directly via a simple closed formula 
obtained using valuation techniques from option-pricing theory. 

This result is far from being obvious. A forward defaultable coupon bond $\overline{P} (t, \tau; {\bf c}, {\bf t})$ 
is the sum of forward defaultable ZCs $\{ \overline{B} (t; \tau, t_i) \}_{i=1\ldots N}$, 
each one following the dynamics (\ref{eq:ForwardDinamicsI}) and then described as a Geometric Brownian Motion (GBM) 
in the case of deterministic volatilities 
(\ref{eq:HW volatility}) we are considering.
No known 
closed formula exists for the running maximum of a sum of GBMs.

In order to get the closed formula, we take the following steps. 
First, we consider a lower and an upper bound of  (\ref{eq:model})
that can be computed via closed formulas. 
Then, we show, calibrating the model parameters for two European issuers, that the difference between the upper 
and lower bounds is negligible for all practical purposes.
We can then use one of the two bounds as the closed-form solution we are looking for; 
in this section we present these bounds and in the next section we show the tightness of their difference.

\bigskip

{\it 
Lower and upper bounds for the {\it sheer liquidity} premium $\Delta_\tau$ (\ref{eq:model}) are:
\be
\sum_{i=1}^N c_i  \overline{B}(t_0, t_i) \left( \pi^L_i (\tau) - \mathbb{P}(t_0, \tau) \right) \le
\Delta_\tau \le
\sum_{i=1}^N c_i  \overline{B}(t_0, t_i) \left( \pi^U_i (\tau) - \mathbb{P}(t_0, \tau) \right)
\label{eq:BoundsFormula}
\en
where the sum is limited to the payment dates $t_i > \tau$ and
\be
\begin{array}{lcl}
\pi^U_i (\tau) & := &
{\displaystyle  
 \frac{4+\Sigma^2_i(\tau)}{2} \, \Phi \left( \frac{\Sigma_i (\tau)}{2} \right)  +
  \frac{\Sigma_i (\tau)}{\sqrt{2 \pi}} \, \exp \left( -\frac{\Sigma_i^2(\tau)}{8} \right) }  \\[4mm]
\pi^L_i (\tau) & := & {\displaystyle \int^1_0 d \eta \frac{e^{-\frac{1}{8}\Sigma^2_N(\tau) }}{\pi \, \sqrt{1 - \eta} \,  \sqrt{\eta}} \,
e^{- \frac{ \eta}{2} \, \Sigma_i(\tau) \, \left(\Sigma_i(\tau) - \Sigma_N(\tau) \right) } }\\[3mm]
& & \; \; {\displaystyle  \left\{ 1 +  
   \sqrt{\frac{\pi \, (1-\eta)}{2}}  \, \Sigma_N(\tau) \, e^{ \frac{1 - \eta}{8} \, \Sigma^2_N(\tau) } \, 
\Phi \left[ \frac{  \sqrt{1-\eta}}{2} \, \Sigma_N(\tau) \right]  \right\} } \\[3mm]
& & \;\; {\displaystyle  \left\{ 1 +  
 \sqrt{\frac{\pi \, \eta}{2}}  \,  \left(  2 \, \Sigma_i(\tau) - \Sigma_N(\tau) \right)  \, e^{ \frac{\eta}{8} \,  \left(  2 \, \Sigma_i(\tau) - \Sigma_N(\tau) \right) ^2 } \, 
 \Phi \left[  \frac{  \sqrt{\eta}}{2} \, \left(  2 \, \Sigma_i(\tau) - \Sigma_N(\tau) \right) \right] 
\right\} } 
\end{array}
\label{eq:pi}
\en
The cumulated volatility is
\[
\Sigma^2_i (\tau) := \displaystyle \int^\tau_{t_0} v^2(s; \tau, t_i) \, ds =
 \zeta_i^2 \, \frac{1 - e^{- 2  \hat{a} \tau  }}{2 \hat{a}}
\]
where $\zeta_i$ is defined in equation (\ref{eq:Volatility}),
$\Phi[\bullet]$ is the standard normal CDF and the issuer survival probability up to the time-to-liquidate is
\be
\label{eq: survProb}
\displaystyle  \mathbb{P}(t_0, \tau) = \exp\left\{ { -\int^\tau_{t_0} \psi_s ds + \frac{\hat{\gamma}^2}{2}  \int^\tau_{t_0}  \overline{\sigma}^2(s, \tau) \, ds} \right\} \; ,
\en
with $ \psi_s $ the deterministic part of the intensity introduced in (\ref{eq: dynamics}) 
and $ \overline{\sigma}(s, \tau)$ is defined in (\ref{eq:HW volatility}).
}


\bigskip

The bounds  for the {\it sheer liquidity} premium $\Delta_\tau$ (\ref{eq:BoundsFormula}) are the key theoretical result of this paper: 
the interested reader can found the deduction of these inequalities in Appendix C.
In Section \ref{sec:NumericalResults} we show for two issuers that these bounds appear to be very tight, 
being their difference on the order of $10^{-8}$ the face value in the worst case (i.e., in Euro terms, it corresponds to \euro $1$ every \euro $100$ million):   
it can be considered negligible for all practical purposes. 

\bigskip

In practice, either of the two closed form solutions (lower or upper bound) can be used 
indifferently, and in particular, the simplest expression of the two bounds, i.e. the upper bound.
This fact allows defining, in an elementary way, a  {\it sheer liquidity} spread as done in the next subsection.

\subsection{The sheer liquidity spread}
\label{subsec:LiquiditySpread}

A consequence of 
equation (\ref{eq:BoundsFormula})
and of the tightness of the difference between the two bounds is that
the illiquid corporate coupon price\footnote{From a notational point view it is standard in the literature to indicate 
a risk-free coupon bond with $P$ and 
a defaultable coupon bond with $\overline{P}$ \citep[see, e.g.][]{Schonbucher}. 
We indicate with $\overline{\overline{P}}_\tau$ the illiquid defaultable coupon bond, 
where also the liquidity risk is considered with a time-to-liquidate $\tau$ in addition to the interest rate and credit risks.}
is
\be
\overline{\overline{P}}_\tau (t_0; {\bf c}, {\bf t}) := \overline{P} (t_0; {\bf c}, {\bf t}) - \Delta_\tau = 
\sum^{N}_{i=1} c_i \overline{B} (t_0, t_i) \, \left( 1 + \mathbb{P}(t_0, \tau) - \pi^U_i (\tau)\right) 
\label{eq:ClosedFormula}
\en
where $\pi^U_i (\tau)$ is defined in (\ref{eq:pi}) and the survival probability
$\mathbb{P}(t_0, \tau)$ can be found in (\ref{eq: survProb}). We can also define an illiquid ZC as
\[
\overline{\overline{B}}_\tau (t_0, t_i) :=  \overline{B} (t_0, t_i) \, \left( 1 + \mathbb{P}(t_0, \tau) - \pi^U_i (\tau)\right) \; \; 
\]
and its {\it sheer liquidity} spread as
\be
L_\tau(t_i) := - \frac{1}{t_i - t_0} \ln \frac{\overline{\overline{B}}_\tau(t_0, t_i)}{\overline{B}(t_0, t_i)} = - \frac{1}{t_i - t_0} \ln \left( 1 + \mathbb{P}(t_0, \tau) - \pi^U_i (\tau) \right) \;\; .
\label{eq:LiquidityBasis}
\en

Thus, we can decompose the illiquid ZC bond into three components: risk-free discount, credit, and liquidity:
\[
\overline{\overline{B}}_\tau (t_0, T) =
\underbrace{ e^{- R(T) (T - t_0)}}_{risk-free} \; 
\underbrace{ e^{- Z(T) (T - t_0)}}_{credit} \; 
\underbrace{ e^{- L_\tau(T) (T - t_0)}}_{liquidity} 
\]
where $R(T)$ is the Zero rate and $Z(T)$ is the Zeta spread of the liquid bond. 
This corresponds to the idea of \citet{Jarrow2001default} 
where liquidity is seen
as a
component in bond yield in addition to the default risk component; it is also 
in line with the practice of market makers in their day-to-day activities: 
they add  to the bond credit spread a basis related to liquidity.

\bigskip

It is useful to underline that the 
{\it sheer liquidity} spread $L_\tau(t_i)$ in (\ref{eq:LiquidityBasis})
 is not affected by the rate component
and  is affected  only slightly by the credit component: this property suits its name ({\it sheer}).
The liquidity component  in the ZC price  $L_\tau(t_i)$  is function of $\left( 1 + \mathbb{P}(t_0, \tau) - \pi^U_i (\tau) \right)$, 
thus it depends mainly on the cumulated volatility $\Sigma_i (\tau)$, because -- as we'll show in the numerical examples --
the quantity $1 - \mathbb{P}(t_0, \tau)$ is two orders of magnitude smaller than $1- \pi^U_i (\tau)$. 

Let us stress a relevant result of the theoretical formulation 
presented in this study. It allows identifying the two key risk factors in the liquidity spread:   
the volatility of the corresponding liquid bond and the 
time-to-liquidate.\footnote{
The cumulated volatility $\Sigma_i (\tau)$  can be seen as the product between the average bond volatility of $\overline{B}(t; \tau, t_i)$ 
in $(t_0, \tau)$  and the squared root of the ttl. Moreover this average bond volatility is close to the volatility $v(t_0; \tau, t_i)$ because in the cases of interest 
$\hat{a} \, \tau \ll 1$ (cf. also equation (\ref{eq:Volatility})).}

\section{An application to the financial sector in the European bond market}
\label{sec:NumericalResults}

In this section we illustrate the impact of illiquidity applying formula (\ref{eq:ClosedFormula}) to obligations with different maturities issued
by two main financial institutions in Europe. We also show that the difference between the upper and lower bounds 
of the illiquidity premium is negligible for all practical 
purposes.\footnote{In this study we consider the two illustrative examples in the financial sector for three main reasons.
First, almost half of the corporate bond market is composed by financial issues. 
Second, most financial institutions present both some very liquid benchmark issues and several illiquid issues intended for some specific customers of clients' portfolio:
these are the illiquid bond we describe in this study.
Finally, corporate bond options are not always frequently traded; 
for financial institutions liquid proxies of these options are often available.} 


\subsection{Calibration of model parameters}
\label{subsec:Calibration}

The two European financial institutions in Europe  that we consider in this study are  BNP Paribas S.A. (hereinafter BNPP) and
Banco Santander S.A. (Santander) on 10 September 2015 (value date).  The settlement date is 
14 September 2015.\footnote{The settlement date is equal to two business days 
after the value date for both the interest rate and credit products in the Euro-zone.}  
At value date, BNPP was rated A and Santander A- according to S\&P.

\bigskip

\begin{table}[h!]
\begin{center}
\begin{tabular}{|c|c|c|c|}
\hline 
maturity & coupon (\%) & clean price \\ 
\hline 
\hline 
27-Nov-2017 & 2.875 & 105.575 \\ 
\hline 
12-Mar-2018 & 1.500 & 102.768 \\ 
\hline 
21-Nov-2018 & 1.375 & 102.555  \\ 
\hline 
28-Jan-2019 & 2.000 & 104.536  \\ 
\hline 
23-Aug-2019 & 2.500 & 106.927  \\ 
\hline 
13-Jan-2021 & 2.250 & 106.083  \\ 
\hline 
24-Oct-2022 & 2.875 & 110.281 \\ 
\hline 
20-May-2024 & 2.375 & 106.007  \\ 
\hline 
\end{tabular}
\end{center}
\caption{\small Clean prices for BNPP liquid bonds. 
Senior unsecured benchmark issues with maturity less than or equal to $10$ years.
Coupons are annual with day-count convention Act/Act. Prices are end-of-day mid-prices 
on 10 September 2015. }
\label{tab: BNP bond data}	
\end{table}

\begin{table}[h!]
\begin{center}
\begin{tabular}{|c|c|c|c|}
\hline 
maturity & coupon (\%) & clean price  \\ 
\hline 
\hline 
27-Mar-2017 & 4.000 & 105.372 \\ 
\hline 
04-Oct-2017 & 4.125 & 107.358 \\ 
\hline 
15-Jan-2018 & 1.750 & 102.766  \\ 
\hline 
20-Apr-2018 & 0.625 & 99.885  \\ 
\hline 
14-Jan-2019 & 2.000 & 103.984 \\ 
\hline 
13-Jan-2020 & 0.875 & 99.500  \\ 
\hline 
24-Jan-2020 & 4.000 & 112.836 \\ 
\hline 
14-Jan-2022 & 1.125 & 98.166  \\ 
\hline 
10-Mar-2025 & 1.125 & 93.261 \\ 
\hline 
\end{tabular}
\end{center}
\caption{\small Clean prices for Santander  
senior unsecured benchmark issues with maturity less than or equal to $10$ years. 
Coupons are annual with day-count convention Act/Act. Prices are end-of-day mid-prices at value date.}
\label{tab: Santander bond data}	
\end{table} 

As discussed in Section 3,
the closed formula for illiquid bond prices, besides  the bond characteristics (maturity, payment dates, coupons, sinking features,  time-of-liquidate, etc...),   includes the observed
i)  zero-rate curve, 
ii) credit spread term-structure for the issuer of interest, and
iii) bond volatility.
These ``ingredients'' can be calibrated with the market data following standard techniques.

\bigskip

First, the risk-free curve we consider is the OIS curve as the market standard; it has been bootstrapped from OIS quoted rates. 
Quotes at value date are provided by Bloomberg. 
The discount curve $B(t_0, T)$ is bootstrapped following the standard procedure;
OIS rates and discount factors
are reported in \cite{BavieraSwaption}. 

\bigskip

Second, in order to construct the Zeta-spread curve, i.e. the (liquid) credit component in the spread,  
we consider all  senior unsecured benchmark issues (i.e. with issue size larger than \euro $500$ million) with maturity less than or equal to $10$ years. 
Coupons are paid annually with the Act/Act day-count convention for all bonds in both sets.
The closing day mid-prices are reported in Tables \ref{tab: BNP bond data} and \ref{tab: Santander bond data}.

For each one of the two issuers, its time-dependent Zeta-spread curve  
\[
 Z(T) := -\frac{1}{T-t_0} \ln \frac{\overline{B}(t_0, T)}{B(t_0, T)} \;\; 
\]
can be bootstrapped from liquid bond invoice prices \citep[see, e.g.][]{Schonbucher}.
Invoice prices are obtained adding the accrual to the clean prices in Tables \ref{tab: BNP bond data} and \ref{tab: Santander bond data}.
We assume a constant Zeta-spread curve up to the maturity of the bond with the lowest maturity 
and we use a linear interpolation rule on Zeta-spread afterwards; the day-count convention for Zeta-spreads is Act/365, as the market standard.

\bigskip

Finally, the volatility parameters ($\hat{a}$, $\hat{\sigma}$ and $\hat{\gamma}$) should be calibrated on options on corporate bonds.
Unfortunately, prices on liquid options on BNPP and Santander bonds are not available in the market at value date.
We consider a proxy in order to calibrate the volatility parameters; 
we notice that at value date both banks are Systemically Important Financial Institutions (SIFI) and belong to the panel of banks contributing to the Euribor rate.
The dynamics of the spread between the Euribor and the OIS curve can be considered a good proxy of the dynamics of the average credit spread for
financial institutions with the above characteristics. As mentioned in \citet{Ruga}, this spread models the risk related to the Euro interbank market,
and default risk is one important component of this interbank risk. Let us underline that we use this proxy to calibrate only volatility parameters,
while credit spreads are calibrated on issuer liquid bond market.

ATM swaptions on Euribor swap rates are very liquid in Europe: we can use these OTC option contracts 
at $t_0$ as a proxy, in order to calibrate  the volatility parameters.
Swaption ATM normal volatilities are provided by Bloomberg; their values in $t_0$ and the
calibration procedure are reported in \cite{BavieraSwaption}.
Calibrated values are  $\hat{a}=12.94\%$, $\hat{\sigma}=1.26 \%$ and $\hat{\gamma} = 0.07\%$. 

\bigskip

In the two cases analyzed, as already mentioned in Section {2}, the correction to include the default risk up to ttl is small.
All survival probabilities $\mathbb{P}(t_0, \tau)$ are close to $1$: 
we report in Table \ref{tab: SurvivalProb} the default probabilities $1-\mathbb{P}(t_0, \tau)$ in the time interval of interest. 
All values are of order $10^{-4}$.

\begin{table}[h!]
\begin{center}
\begin{tabular}{c|c c}
& {\it BNPP}  & {\it Santander} \\ 
\hline 
2w & $1.27 \times 10^{-4}$ & $1.80 \times 10^{-4}$ \\ 
2m & $5.42 \times 10^{-4}$ & $7.73 \times 10^{-4}$ \\ 
\end{tabular}
\end{center}
\caption{\small Default probabilities $1-\mathbb{P}(t_0, \tau)$ for  BNPP and Santander  for the two ttl of $2$ weeks and $2$ months.}
\label{tab: SurvivalProb}	
\end{table} 

The correction due to the default risk up to ttl is negligible in the liquidity spread:
this fact justifies the decomposition of the bond spread in the three components {\it risk-free}, {\it credit} and {\it liquidity} proposed in 
Section {\bf \ref{subsec:LiquiditySpread}} and the adjective ({\it sheer}) of the liquidity spread we consider.

\subsection{Illiquid bond prices}
\label{subsec:IliiquidPrices}

\begin{figure}
  	\begin{center}
      	\includegraphics[width=0.65\textwidth]{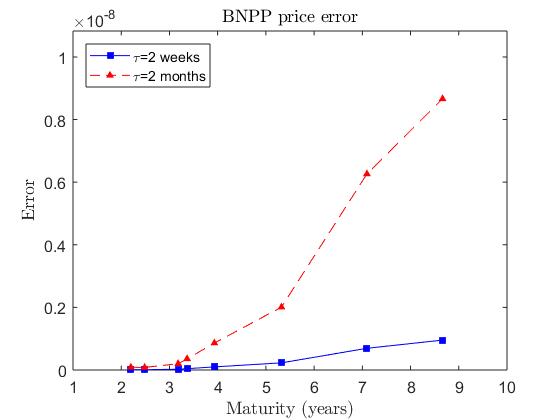}
	\end{center}
  	\vspace{-15pt}
  	\caption{\small Difference between the upper and lower bounds for the {\it sheer liquidity} premium $\Delta_\tau$ for BNPP bonds. 
We consider illiquid bonds with the same characteristics (e.g. coupons, payment dates) as the bonds in Table \ref{tab: BNP bond data} 
with ttl equal to two weeks (continuous blue line and squares) and two months (dashed red line and triangles).  
This difference is on the order of $10^{-8}$ times the face value in the worst-case, and so it is negligible for all practical purposes.}
\label{fig:BNPBounds}
\end{figure}

\begin{figure}
  	\begin{center}
      	\includegraphics[width=0.65\textwidth]{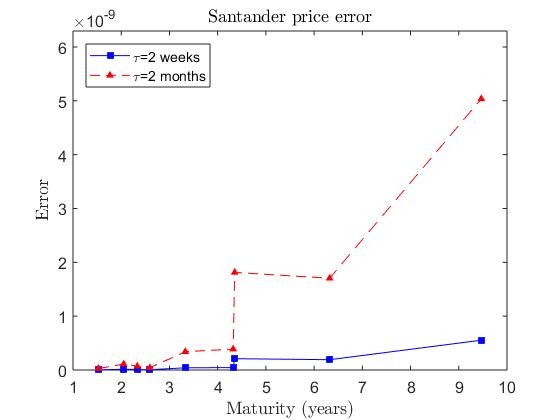}
	\end{center}
  	\vspace{-15pt}
  	\caption{\small Difference between the upper and lower bounds for the {\it sheer liquidity} premium $\Delta_\tau$ for Santander bonds. 
We consider illiquid bonds with the same characteristics (e.g. coupons, payment dates) as the bonds in Table \ref{tab: Santander bond data} 
with ttl equal to two weeks
(continuous blue line and squares) and two months (dashed red line and triangles). This difference is on the order of $10^{-9}$ times the face value.
}
\label{fig:SantanderBounds}
\end{figure}

In this section we show that, 
considering two sets of illiquid bonds with the same characteristics as the liquid bonds
(e.g. coupons and payments dates) and ttl equal to either two weeks or two months, 
the difference between the lower and upper bounds for the {\it sheer liquidity} premium $\Delta_\tau$  is on the order of $10^{-8}$ times
the face value. 
Figure \ref{fig:BNPBounds}  presents this difference for BNPP, and Figure \ref{fig:SantanderBounds} for Santander.
This difference is the maximum error we make if we
evaluate $\Delta_\tau$ with one of these bounds. 
It is negligible for all practical purposes.

Moreover, as a robustness test, we have considered this difference between the two bounds 
for a wide range of volatility parameters around the estimated values, keeping equal all other bond characteristics: 
$\hat{a} \in (0, 30\%)$, $\hat{\sigma} \in (0, 4\%)$ and $\hat{\gamma} \in (0, 0.2\%)$. 
We observe that
this difference is, in the worst case, 
less than $1$ Euro for every million of 
face value. 
Again, we find that 
the difference between the two bounds is negligible for all practical purposes.
This fact allows us to consider indifferently either the lower or the upper bound in (\ref{eq:pi}) as a closed-form solution for $\Delta_\tau$.

\bigskip

\begin{figure}
  	\begin{center}
      	\includegraphics[width=0.65\textwidth]{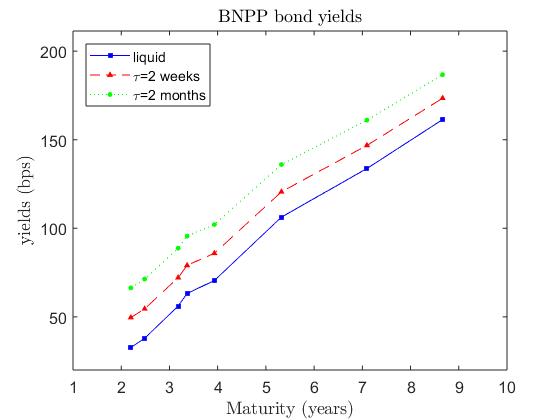}
	\end{center}
  	\vspace{-15pt}
  	\caption{\small BNPP bond yields. We consider all benchmark issues with maturity less than 10y described in Table \ref{tab: BNP bond data} and
their yields (continuous blue line  and squares). We show also the yield obtained 
for illiquid bonds with the same characteristics (e.g. coupons, payment dates) with ttl equal to two weeks 
(dashed red line and triangles) and two months (dotted green line and circles).}
\label{fig:BNPLiquidity}
\end{figure}

\begin{figure}
  	\begin{center}
      	\includegraphics[width=0.65\textwidth]{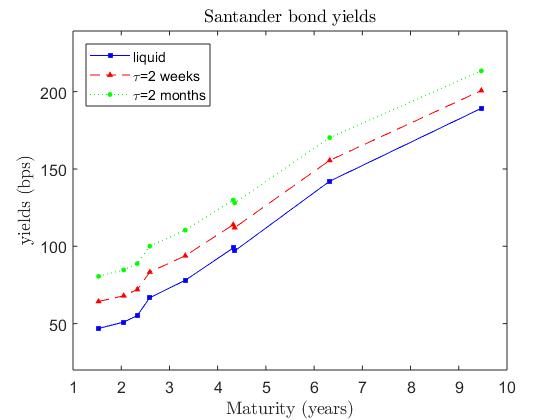}
	\end{center}
  	\vspace{-15pt}
  	\caption{\small Santander bond yields. We consider all benchmark issues with maturity lower than 10y in Table \ref{tab: Santander bond data} and
their yields (continuous blue line  and squares). 
We show also the yield obtained 
for illiquid bonds with the same characteristics (e.g. coupons, payment dates) with ttl equal to two weeks (dashed red line and triangles) and  
ttl equal to two months (dotted green line and circles).}
\label{fig:SantanderLiquidity}
\end{figure}

In Section $3$ we have shown that a {\it sheer liquidity} spread (\ref{eq:LiquidityBasis}) could be added to each ZC in order to take into account liquidity.  
Practitioners often consider a liquidity yield spread as the term that should be added to the yield in order to obtain the illiquid bond price (\ref{eq:ClosedFormula})
\[
\overline{\overline{P}}_\tau (t_0, T; {\bf c}, {\bf t}) =:  \sum^N_{i=1} c_i \; 
e^{- \left[ {\cal Y} (T)+ {\cal L}_\tau(T) \right] \, (t_i - t_0)}
\]
where ${\cal Y}(T)$ is the yield of the corresponding liquid bond $\overline{P} (t_0; {\bf c}, {\bf t})$ and
${\cal L}_\tau(T)$ the  liquidity yield spread for ttl equal to $\tau$.

In Figures \ref{fig:BNPLiquidity} and  
\ref{fig:SantanderLiquidity} we show the liquidity yield spread for BNPP and Santander 
for different bond maturities and ttl equal to two weeks and two months.
We observe that, for the same time-to-liquidate $\tau$, the liquidity yield spread ${\cal L}_\tau(T)$ depends only slightly on bond maturity $T$.


\section{Conclusions}
\label{sec:Conclusions}

In this paper we have proposed an approach via a reduced-form model for pricing illiquid corporate bonds 
when the corresponding liquid bonds are observed in the market. 
It allows a closed formula (\ref{eq:ClosedFormula}) for illiquid bond prices when they are not available in the marketplace
and a direct calibration of parameters on the risk-free curve, the Zeta-spread curve of the issuer of interest and its bond volatility.
We have shown a detailed model calibration for two European corporate issuers in the financial sector. 

Our approach models the difference between liquid and illiquid coupon bond prices, named {\it sheer liquidity} premium, as a right in investor hands.
The formula is deduced in two steps: i) bounding from above and below the {\it sheer liquidity} premium (\ref{eq:model}) in a DHJM (\ref{eq: dynamics}) and
ii) showing the equivalence of 
these two bounds for all practical purposes.

\bigskip

This closed formula (\ref{eq:ClosedFormula})  is simple and allows to identify the two key drivers of the {\it sheer liquidity}:
the bond volatility and 
the ``time-to-liquidate'' a given bond position. 
It can be used by practitioners for different possible applications. Let us mention some of them.

This model can support market makers in their day-to-day activities.
On the one hand, the ttl
parameter can be evaluated {\it ex ante} by an {\it experienced trader} with a deep knowledge of the characteristics 
of that particular illiquid market 
(concentration, frequency for trades with similar characteristics observed in the recent past) 
who desires to liquidate a given position; the formula gives a theoretical background for the market practice of adding a liquidity spread to the bond yields 
either when pricing illiquid issues or when receiving them as collateral.
On the other hand,
the formula can also be used to obtain an ``implied time-to-liquidate'' from market quotes if both liquid and illiquid prices are available, 
translating observable spreads into a time lag for liquidating a position and hence providing an interesting piece of information for market 
participants.\footnote{The idea of measuring the implied time-to-liquidate has been first suggested by \citet{AbudyRaviv}.}

Moreover, the model can be useful also to risk managers. 
By offering an explicit relationship between the bond volatility and the {\it sheer liquidity} premium, 
it  offers to risk managers a way to justify the market practice of setting
limits on illiquid bond positions based on the volatility of similar liquid bonds
and it gives a theoretical background for 
haircuts of 
illiquid bonds accepted as collateral.
Moreover, the ttl can be backtested {\it ex post} by risk managers, who, thanks to the transaction data made recently available, 
can measure the average time needed for liquidating a position in an illiquid corporate bond of a given size. 


\bigskip
The proposed approach clarifies that illiquidity is an intrinsic component of the bond spread 
mainly related to the cumulated volatility. In the presence of a liquid credit curve, 
it allows to disentangle the two components, credit and liquidity, in the observed spread over the risk free rate.

\section*{Acknowledgments}

We thank all participants  to the seminar at the European Investment Bank (EIB) and
conference participants 
to the $8^{th}$ General AMaMeF Conference  in Amsterdam, 
to the $18^{th}$ Workshop on Quantitative Finance, 
to the Vienna Congress on Mathematical Finance, 
to the SIAM Conference on Financial Mathematics \& Engineering in Toronto and  %
to the $10^{th}$ Annual Financial Market Liquidity Conference in Budapest. 
We are grateful in particular to  Pino Caccamo, Robert Czech, Jose M. Corcuera,  
Damir Filipovic, Szabolcs Gaal, D\'aniel Havran, 
Fabrizio Lillo, Jan Palczewski, Andrea Pallavicini, Oleg Reichmann, Hans Schumacher,
Pierre Tychon and Niklas Wagner for useful comments. 

R.B. acknowledges EIB financial support under the EIB Institute Knowledge Programme.
The findings, interpretations and conclusions presented in this document are entirely those of the authors and should not be attributed in any manner to the EIB.
Any errors remain those of the authors. 

\bibliography{illiquidity}
\bibliographystyle{tandfx}

\section*{Notation and Shorthands}

\bigskip

{\bf Shorthands}

\[
\begin{array}{lcl}
{\rm CDF} & : & {\rm Cumulative \; Distribution \; Function}   \\[1mm]
{\rm cf.} & : & {\rm compare; \; from \; Latin: \; confer}\\[1mm]
{\rm FR} & : & \text{fractional-recovery (model)} \\[1mm] 
{\rm DHJM} & : & \text{Defaultable  HJM framework} \\[1mm] 
{\rm GBM} & : & \text{Geometric Brownian Motion} \\[1mm]
{\rm IPO} & : & \text{Initial Public Offering}  \\[1mm] 
{\rm MiFID} & : & \text{Markets in Financial Instruments Directive}    \\[1mm] 
{\rm OTC} & : & \text{Over The Counter}  \\[1mm] 
{\rm OU} & : & \text{Ornstein--Uhlenbeck process}    \\[1mm] 
{\rm pdf} & : & {\rm probability \; density \; function}   \\[1mm]
{\rm s.t.} & : & {\rm such \; that}   \\[1mm]
{\rm ttl} & : &  \text{time-to-liquidate}    \\[1mm]
{\rm w.r.t.} & : & {\rm with \; respect \;  to} \\[1mm] 
{\rm TRACE} & : & \text{Trade Reporting and Compliance Engine}    \\[1mm]
{\rm ZC} & : & \text{zero-coupon bond} \\[1mm] 
\end{array}
\; \; .
\]

\newpage

{\bf Notation}
\begin{center}
\begin{tabular}{|l|l|} \hline
{\bf Symbol} & {\bf Description} \\
\hline
$ {\hat a},{\hat \sigma},  {\hat \gamma}$ & parameters in short rate $r_t$ and intensity $\lambda_t$ dynamics (\ref{eq: dynamics}) \\
$B(t,T)$ & risk-free zero-coupon (ZC) bond at $t$ with maturity $T$\\
${\overline B}(t,T)$ & defaultable ZC bond at $t$ with maturity $T$ and zero-recovery\\
$\overline{\overline{B}}_\tau(t, T)$ & illiquid defaultable ZC bond at $t$ with maturity $T$ , zero-recovery and ttl equal to $\tau$ \\
${\overline B}(t;\tau,T)$ & forward ZC bond\\
$ {\bf c} = \{ c_i\}_{i=1,\ldots,N}$ & defaultable  coupon bond flows (coupons and face value) \\
$\overline{P}(t_0; {\bf c}, {\bf t})$ & defaultable coupon bond at $t_0$  \\
$\overline{\overline{P}}_\tau (t_0; {\bf c}, {\bf t})$ & illiquid defaultable coupon bond at $t_0$ with ttl equal to $\tau$\\
$\overline{P}(t,\tau; {\bf c}, {\bf t})$ & forward defaultable  coupon bond at $t$, paid at $\tau$ \\
$\mathbb{P}(t_0, \tau)$ & corporate issuer survival probability up to the time-to-liquidate $\tau$ \\ 
$\Delta_{\tau}$ &  {\it sheer liquidity} premium with ttl equal to $\tau$\\
${D}(t,T)$ & stochastic discount factor, equal to $\displaystyle \exp \left( - \int^T_{t} r_s \, ds \right) $\\
${\overline{D}}(t,T)$ & defaultable stochastic discount factor, equal to $\displaystyle \exp \left(  - \int^T_{t} \overline{r}_s \, ds \right) $\\
$\mathbb{E}[\bullet] \; \& \; \mathbb{E}^{(\overline{\tau})}[\bullet]$ & expectation under the risk neutral \& under the $\tau$-defaultable-forward measure\\
$L_\tau (t_i)$ & {\it sheer liquidity} spread for a ZC with maturity $t_i$ and  ttl $\tau$ \\
${\cal L}_\tau (T)$ & liquidity yield spread for a coupon bond with maturity $T$ and  ttl $\tau$\\
$N$ & number of coupons in the defaultable bond   \\
${\cal N}_t$ & Cox process with stochastic intensity $\lambda_t$ that models default of the corporate issuer \\
$q$ & loss fraction given default in FR models; $q = 1^-$  reproduces the zero-recovery case \\
$\lambda_t$ & stochastic intensity at time $t$\\  
$r_t$ & risk-free short rate at time $t$\\  
$\overline{r}_t$ & defaultable short rate at time $t$, defined as $r_t + q \lambda_t$ \\  
$\rho$ & instantaneous correlation matrix in $\Re^{d \times d}$ s.t. $dW^{(i)}_t \; dW^{(j)}_t = \rho_{i\, j} \, dt$ \\
$\sigma(t,T) \,\&\, \overline{\sigma}(t,T)$ & DHJM risk-free and  defaultable ZC volatilities between $t$ and $T$ in $\Re^d$ \\
$\Sigma_i(\tau)$ & cumulated volatility s.t. $ \Sigma^2_i (\tau) := \displaystyle \int^\tau_{t_0}  v^2(s; \tau, t_i) \, ds $ \\
$t_0$ & value date ($t_0 =0$) \\
$t_d$  & time to default \\
$\tau$ & time-to-liquidate (ttl)  \\
$ {\bf t} = \{ t_i\}_{i=1,\ldots,N}$ & payment dates  of the defaultable  coupon bond with maturity $t_N \equiv T$\\
$v(t; \tau, T)$ & equal to $\sigma(t,T) - \sigma(t,\tau) $ \\ 
$W_t$ & vector of correlated Brownian motions in $\Re^d$ s.t. $dW^{(i)}_t \; dW^{(j)}_t = \rho_{i\, j} \, dt$ \\
$x \cdot y$ & scalar product between $x,y \in \Re^d$ \\
$x^2$ & an abbreviation for scalar product $x \cdot \rho x$ with $x \in \Re^d$ and $ \rho \in \Re^{d \times d}$ correlation \\
${\cal Y}(T)$ & yield of the corporate bond $\overline{P} (t_0; {\bf c}, {\bf t})$ with maturity $T$ \\
\hline
\end{tabular}
\end{center}

\bigskip


\section*{Appendix A}

In this Appendix we recall some basic properties of DHJM with fractional recovery \citep[see, e.g.][]{DuffieSingleton1999, Schonbucher1998}.
We also show an application of these properties to the price of illiquidity (\ref{eq:model}).

\bigskip

Absence of arbitrage require that
the instantaneous risk-free rate $r_t$ and the defaultable one $\overline{r}_t $ satisfy
\be
\left\{
\begin{array}{rcl}
r_t &:=& {\displaystyle - \frac{\partial \ln B(t_0,t)}{\partial t} + 
\frac{1}{2} \int^t_{t_0} \frac{ \partial }{ \partial t } 
\sigma(s,t)^2 \; ds -
\int^t_{t_0} \frac{ \partial }{ \partial t } \sigma(s,t) \cdot dW_s} \\[4mm]
\overline{r}_t &:=& {\displaystyle - \frac{\partial \ln \overline{B}(t_0,t)}{\partial t} + 
\frac{1}{2} \int^t_{t_0} \frac{ \partial }{ \partial t } 
\overline{\sigma}(s,t)^2 \; ds -
\int^t_{t_0} \frac{ \partial }{ \partial t } \overline{\sigma}(s,t) \cdot dW_s } \; ,
\end{array}
\right.
\label{eq:rates DHJM}
\en
that correspond to equations (25) and (17) in \citep{Schonbucher1998}, with $\overline{B}(t_0,t)= \overline{B}_{q}(t_0,t) \; \forall t$.

\bigskip

The above dynamics for $\overline{r}_t$ implies that, at value date $t_0$, the relation between defaultable discount and defaultable ZC is
\be
\displaystyle \overline{D}_q(t_0, \tau) = \overline{B}(t_0, \tau)  \exp \left\{ 
- \frac{1}{2} \int^\tau_{t_0} \overline{{\sigma}}^2 (s,\tau) \; ds +
\int^\tau_{t_0} \overline{{\sigma}} (s,\tau)  \cdot  d {W}_s \right\} \; .
\label{eq:Lemma2}
\en
This relation is the same of the one in the HJM for risk-free rates, because all quantities are continuous \citep[see, e.g.][]{Musiela}.
A consequence of equation (\ref{eq:Lemma2}) is that  the $\overline{\tau}$-forward measure presents 
an interesting property at value date $t_0=0$
\be
\mathbb{E} \left[   \overline{D}_q(t_0, \tau) \; \bullet \; | {\cal G}_{0}  \right] = \overline{B}(t_0, \tau) \; \mathbb{E}^{(\overline{\tau})} \left[ \; \bullet \; | {\cal G}_{0} \right]
\label{eq:ChangeMeasProperty}
\en
that is an application of Girsanov's theorem \citep[see, e.g.][]{Musiela}.

\bigskip 

Moreover, from equations (\ref{eq:DHJM}) 
we get that the value of the defaultable ZC at a generic time $t$ starting from the initial condition in $t_0$ is
\[
\overline{B}(t,T) = \overline{B}(t_0,T) \; (1-q)^{{\cal N}_t} \;
\exp \left\{ 
\int^t_{t_0} \left[ \overline{r}_s - \frac{1}{2} \overline{{\sigma}}^2 (s,T) \right] ds +
\int^t_{t_0} \overline{{\sigma}} (s,T) \cdot d {W}_s \right\} \; .
\]
which can be obtained  using the Generalized It\^o lemma \citep[see, e.g.][eq.(79) p.185]{Schonbucher1998}.
Thus, the default of $\overline{B}_q(t, T)$ occurs when the process jumps. In case of a jump at time $t$, the jump size is 
\[
\Delta \overline{B}_q(t, T) = \overline{B}_q(t, T) - \overline{B}_q(t^-, T) = - \overline{B}_q(t^-, T) \, q \, d {\cal N}_t
\]
i.e. the ZC loses a fraction $q$ of its pre-default value, as indicated when introducing the model in (\ref{eq:FR main}).

\bigskip 

Finally, we show that, within the DHJM framework with $q=1^-$, the price of illiquidity (\ref{eq:model})  can be simplified
\be
\label{eq: lemma3Appendix}
\begin{split}
\Delta_\tau 
 & = 
\mathbb{E} \left[ \overline{D} (t_0, \tau) M_\tau | {\cal G}_{0} \right] -
\mathbb{E} \left[ \overline{D} (t_0, \tau) \overline{P}(\tau; {\bf c}, {\bf t}) | {\cal G}_{0} \right] + \left( 1- \mathbb{P}(t_0, \tau) \right) \overline{P}(t_0; {\bf c}, {\bf t}) = \\
& = \overline{B} (t_0, \tau) \left\{
\mathbb{E}^{(\overline{\tau})} \left[ M_\tau | {\cal G}_{0} \right] -
\mathbb{E}^{(\overline{\tau})} \left[ \overline{P}(\tau; {\bf c}, {\bf t}) | {\cal G}_{0} \right] \right\} + \left( 1- \mathbb{P}(t_0, \tau) \right) \overline{P}(t_0; {\bf c}, {\bf t}) = \\
& = \overline{B} (t_0, \tau) \,
\mathbb{E}^{(\overline{\tau})} \left[ M_\tau | {\cal G}_{0} \right] -
\mathbb{P}(t_0, \tau) \, \overline{P}(t_0; {\bf c}, {\bf t}) \;\; ,
\end{split}
\en 
where $\mathbb{P}(t_0, \tau)$ is the issuer survival probability up to the time-to-liquidate.
The first line comes from iterated expectation,
the second line is due to the change of measure property in the $\overline{\tau}$-forward measure (\ref{eq:ChangeMeasProperty}),   
while the last equality follows because a coupon bond and the corresponding forward are related via  
$\overline{P}(t_0; {\bf c}, {\bf t}) = \overline{B} (t_0, \tau) \, \overline{P}(t_0, \tau; {\bf c}, {\bf t})$, that is equivalent to (\ref{eq:ForwardCouponBond}).

\section*{Appendix B}

In this Appendix we 
deduce two properties, useful in the derivation of the closed form bounds of {\it sheer liquidity} premium:
the inequalities (\ref{eq:BasicInequalities}) and the joint probability (\ref{eq:JointProbability}).

\smallskip

First, the following inequalities hold: 
\be
\sum_i 
c_i  \overline{B}(t^*;\tau,t_i) \le
\max_{t \in [t_0, \tau]} \left\{  \sum_i 
c_i  \overline{B}(t;\tau,t_i) \right\} \le
\sum_i 
c_i \max_{t \in [t_0, \tau]}   \overline{B}(t;\tau,t_i)  \;\; \forall t^* \in  [t_0, \tau], \; t_i \ge \tau
\label{eq:BasicInequalities}
\en
where the sum over $i$ is limited to all coupons with payment date $t_i$ larger than $\tau$.

The left inequality is obvious since the maximum value of a function on the time interval $[t_0, \tau]$ is greater than the same function's values at 
any other time $t^*$ in the interval. The right inequality is due to the fact that the maximum of a sum is less than or equal to the sum of the maxima. 

\smallskip

In particular we can choose $t^*$ 
equal to the time-location 
\be
t^* = \min \left\{ t' \, \left| \, \overline{B}(t';\tau,t_N) = \max_{t \in [t_0, \tau]}  \overline{B}(t;\tau,t_N) \right. \right\} \; \; ,
\label{eq:optimal time}
\en
i.e. equal to the (first) time when the last forward ZC, $ \overline{B}(t;\tau,t_N) $, reaches its maximum in the interval $[t_0, \tau]$.

\bigskip

Second, given $x(t) := c\, t + W_t$ a 1-dimensional Wiener process with drift $c \; t$ where $c \in \Re$,
the joint probability of i) the maximum $y := \max [ x (t); t \in (0,T) ] $ and ii) its time location $\theta \in (0,T)$, is   
\be
p(\theta, y; c, T) = \frac{1}{\pi}\,\frac{y}{\sqrt{T - \theta} \,  \theta^{3/2}}\, e^{-\frac{c^2 T}{2} -\frac{y^2}{2 \theta} + c y } \,
\left\{ 1 -  
 \sqrt{2 \pi \,( T - \theta)} \, c \,  e^{ \frac{c^2 (T- \theta)}{2}} \,  
 \Phi \left[ - c \, \sqrt{T - \theta} \right] 
\right\}
\label{eq:JointProbability}
\en
with $y = x(\theta) >0$. 

This  joint probability $p(\theta, y; c, T)$ (\ref{eq:JointProbability})  is due to a known result in \citet{Shepp}.
Consider the density $p(\theta, y, x; c, \sigma^2)$ in equation (1.5) in  \citet[][p.424]{Shepp}, where $x$ is the endpoint $x(T)$.
The joint probability $p(\theta, y; c, T) $ is obtained by setting $\sigma =1$ and by integrating over $x\in (-\infty, y) $. 

\section*{Appendix C}

In this Appendix we deduce (\ref{eq:BoundsFormula}), 
the main theoretical result of the paper, where we show that 
there exit a lower and an upper bound for the {\it sheer liquidity} premium that can be expressed in a simple closed form.

\smallskip

The upper bound in (\ref{eq:BoundsFormula}) is obvious given equation (\ref{eq:BasicInequalities}) and after observing that each ZC $ \overline{B}(t; \tau, t_i)$ in equation (\ref{eq:ZCdynamycs}) 
follows a driftless GBM with volatility $v(t; \tau, t_i)$ under the  $\overline{\tau}$-forward measure. 
Thus,  the expected value of the running maximum of the $i^{th}$ driftless GBM  $\overline{B}(t; \tau, t_i)$ for $t \in[t_0, \tau]$ 
takes the form $\overline{B}(t_0; \tau, t_i) \, \pi^U_i (\tau)$ \citep[see, e.g.][and references therein]{Longstaff}.

\bigskip

The lower bound  in (\ref{eq:BoundsFormula}), according to the same equation (\ref{eq:BasicInequalities}), 
is the sum over $i$ of the expected values of $ \overline{B}(t^*; \tau, t_i)$, with $i=1\ldots N$.
They are computed at time $t^*$ s.t.
$ \overline{B}(t^*; \tau, t_N)$ reaches its first maximum (for a given realization of the process).
In lower bound case, we can define $\pi^L_i (\tau)$ s.t.
\[
\overline{B}(t_0; \tau, t_i) \, \pi^L_i (\tau) :=
{\mathbb E}^{(\overline{\tau})} \left\{ \overline{B}(t^*; \tau, t_i) \right\} \;\; .
\]
Using the separability property of the volatility (\ref{eq:Volatility}), we get
\[
{\mathbb E}^{(\overline{\tau})} \left\{ \overline{B}(t^*; \tau, t_i) \right\} = 
\overline{B}(t_0; \tau, t_i) \, {\mathbb E}^{(\overline{\tau})} 
\left\{ \exp \left[ \zeta_i \left( - \frac{1}{2} \zeta_i \int^{t^*}_{t_0} \nu^2 (s) \, ds + \int^{t^*}_{t_0} \nu (s) \, dW^{(\overline{\tau})}_s \right) \right] \right\} \; .
\]
By means of the change of time 
\[
\tilde{t} := \tilde{t}(t) := \int^{t}_{t_0} \nu^2 (s) \, ds \; \in \, (0, \tilde{\tau})
\]
where $\tilde{\tau}$ stands for $\tilde{t}(\tau)$,
we get $ d W^{(\overline{\tau})}_{\tilde{t}} = \nu (t) \, dW^{(\overline{\tau})}_t $
and 
\[
\begin{split}
{\mathbb E}^{(\overline{\tau})} \left\{ \overline{B}(t^*; \tau, t_i) \right\} & = 
\overline{B}(t_0; \tau, t_i) \, {\mathbb E}^{(\overline{\tau})} 
\left\{ \exp \left[ \zeta_i \left( - \frac{1}{2} \zeta_i  \, \theta + W^{(\overline{\tau})}_\theta \right) \right] \right\} \\
& =
\overline{B}(t_0; \tau, t_i) \, {\mathbb E}^{(\overline{\tau})} 
\left\{ \exp \left[ \zeta_i \left( - \frac{1}{2} (\zeta_i -  \zeta_N) \, \theta + x(\theta) \right) \right] \right\}
\end{split}
\]
where we have defined
the drifted Brownian motion $x(\tilde{t})  := - \zeta_N\, \tilde{t} /2  + W^{(\overline{\tau})}_{\tilde{t}}$ with  $\tilde{t} \in (0, \tilde{\tau})$.
We also define
 $y:=x(\theta)$ its (first) maximum value with
\[
\theta := \tilde{t}(t^*)   \; \; .
\]

Using equation (\ref{eq:JointProbability}), let us observe that
\[
{\displaystyle {\mathbb E}^{(\overline{\tau})} 
\left\{ \exp \left[ \zeta_i \left( - \frac{1}{2} (\zeta_i -  \zeta_N) \, \theta + x(\theta) \right) \right] \right\} =
\int^{\tilde{\tau}}_0 d \theta \int^{+ \infty}_0 dy \, p \left(\theta, y; - \frac{\zeta_N}{2}, {\tilde{\tau}} \right)  
e^{\zeta_i \left( - \frac{1}{2} (\zeta_i -  \zeta_N) \, \theta + y \right)} }
\]
where $p(\theta, y; - \zeta_N/2, {\tilde{\tau}} ) $ has been obtained in (\ref{eq:JointProbability}). 
After computing the integral w.r.t. $y$ we get $\pi^L_i (\tau) $ in the lower bound.
That proves the proposed lower and upper bounds  (\ref{eq:BoundsFormula}) for the {\it sheer liquidity} premium.

\bigskip

Moreover, the survival probability (\ref{eq: survProb}) can be computed for model (\ref{eq: dynamics}) starting from its definition
\[
\mathbb{P} (t_0, \tau) := 
\mathbb{E} \left[ \mathbbm{1}_{t_d > \tau} | {\cal F}_{0} \right] =
\mathbb{E} \left[ \exp - \left( \int^\tau_{t_0} \lambda_t dt \right)  | {\cal G}_{0} \right]
\]
with $ \lambda_t$ given by equation (\ref{eq: dynamics}) and $x_t$ is the solution of the OU process with zero mean
\[
x_t =  \int^\tau_{t_0} dW_t \, \overline{\sigma}(t,\tau) \;\; ,
\]
where $ \overline{\sigma}(s, \tau)$ is defined in (\ref{eq:HW volatility}).
We get
\[
\mathbb{P} (t_0, \tau) = 
\exp \left(- \int^\tau_{t_0} \psi_t \, dt \right)  
\mathbb{E} \left[ \exp \left( - {\hat \gamma} \int^\tau_{t_0} dW_t \, \overline{\sigma}(t,\tau)  \right) | {\cal G}_{0} \right] =
\exp  \int^\tau_{t_0} dt \, \left( - \psi_t + \frac{{\hat \gamma}^2}{2} \overline{\sigma}(t,\tau)^2 \right) \;\; .
\]

  
\bigskip

Finally,
it can be interesting to comment on the reason
why lower and upper bound are very close in practice.

The lower bound is computed on the time $t^*$ in (\ref{eq:optimal time}) that maximizes the forward ZC with expiry in $t_N$ (with the face value),
while the upper bound is the sum of the running maximum 
of each forward ZC $ \overline{B}(t; \tau, t_i)$, with $i=1\ldots N$.

\smallskip

On the one hand,  
a forward coupon bond (\ref{eq:ForwardCouponBond}) 
is the sum of forward ZCs which have different weights $c_i$ 
with the last one $c_N$ (that contains the face value) generally
two orders of magnitude larger than the others. 

On the other hand, 
each $i^{th}$ forward ZC, in the forward defaultable coupon bond $\overline{P} (t, \tau ; {\bf c}, {\bf t})$, follows a GBM whose
maximum over time takes place at 
\[
 t^{max}_i := \argmax_{t\in[t_0, \tau]} \left[ - \frac{\zeta_i}{2} \int^t_{t_0} \nu^2 (s) \, ds + \int^t_{t_0} \nu (s) \, dW^{(\overline{\tau})} (s) \right] \;.
\]
In the above expression, the stochastic part is exactly the same $\forall i$ and differs only for the $\zeta_i$ term in the 
deterministic drift part, where $\zeta_i$ is small, once one considers the parameters calibrated with market data.
For this reason, 
the time-location of the maximum is exactly the same for all forward ZCs and equals $t^*$ in (\ref{eq:optimal time}) in most scenarios.



\end{document}